\documentclass[12pt, a4paper] {article}

\usepackage{amsfonts}
\usepackage{verbatim}
\usepackage{color}
\usepackage{amsmath}
\usepackage{graphicx}
\usepackage{bm}
\usepackage{amssymb}
\usepackage{latexsym}
\usepackage{mathrsfs}
\usepackage{textcomp}
\usepackage{titlesec}
\usepackage{ulem}
\titleformat*{\section}{\normalfont\bfseries}
\titleformat*{\subsection}{\normalfont\itshape}

\begin{document}

\begin{center}
{\bf\large{Revisiting novel symmetries in coupled $\mathcal{N}$ = 2 supersymmetric quantum systems:
examples and supervariable approach} }

\vskip 1.5 cm

{\sf{ \bf Aditi Pradeep\footnote{Current affiliation: Department of Physics and Astronomy, University of British Columbia,
Vancouver Campus, 325 - 6224 Agricultural Road, Vancouver, BC V6T 1Z1, Canada}, Anjali S, Binu M Nair and Saurabh Gupta}}\\
\vskip .1cm
{\it Department of Physics, National Institute of Technology Calicut,\\ Kozhikode - 673 601, Kerala, India}\\
\vskip .15cm
{E-mails: {\tt aditipradeep314@gmail.com, anjalisujatha28@gmail.com, binu5120@gmail.com, saurabh@nitc.ac.in}}

\end{center}

\vskip 1 cm

\noindent
{\bf Abstract:}
 We revisit the novel symmetries in $\mathcal{N}$ = 2 supersymmetric (SUSY) quantum mechanical (QM) models
by considering specific examples of coupled systems. Further, we extend our analysis to a general case and list out all
the novel symmetries. In each case, we show the existence of two sets of discrete symmetries that correspond to
the Hodge duality operator of differential geometry. Thus, we are able to provide a proof of the conjecture which points
out the existence of more than one set of discrete symmetry transformations corresponding to the Hodge duality operator. Moreover, we derive
on-shell nilpotent symmetries for a generalized superpotential within the framework of supervariable approach.

\vskip 1.5 cm



\newpage

\section {Introduction}
Particle interactions in nature have always been a question of immense scientific
interest. The quest to unify the four fundamental interactions, namely; gravitational, electromagnetic, strong and weak has drawn significant attention over the past century. However, the consistent unification of gravity with strong and electroweak
interactions is still a profound task of interest \cite{franck,11,wess}. This led to the birth of a popular principle known as
supersymmetry \cite{3}. It is one of the most beautiful principles which has rich background support from physics as well as
mathematics. It unifies two different classes of particles - fermions and bosons \cite{1,nilles,sohn}  - with mathematical
substantiality encompassed within a graded Lie algebra which uses anti-commutation relations
instead of commutation relations, thereby enabling the unification of fundamental interactions \cite{fayet, 2}.

The concept of supersymmetry has various applications beyond grand unification. In non-relativistic
quantum mechanics, supersymmetry has paved the way for an elegant factorization technique
of the Hamiltonian. This allows the calculation of energy spectrum of a given
Hamiltonian without resorting to more complicated methods of solving the corresponding Schr\"{o}dinger equation
(cf. \cite{3} for details).

 The SUSY harmonic oscillator has been regarded as one of the simplest examples as far as \(\mathcal{N}\) = 2 SUSY QM models are concerned (see e.g. \cite{18,5,6,7}). In the recent past, the correspondence among the continuous symmetries
of SUSY harmonic oscillator with the de Rham cohomological operators of differential geometry has been established at the algebraic level \cite{9,8} and it has been shown that the discrete symmetries correspond to the Hodge duality operator of differential geometry \cite{12,13,14}. Further a detailed note about the existence of discrete symmetries in any general \(\mathcal{N}\) = 2 SUSY QM model has also been provided and it has been shown that they provide an example for Hodge theory \cite{10}. It is worthwhile to mention that similar kind of one-to-one correspondence among the symmetry transformations and de Rham cohomological operators has already been shown to exist in some of the field theoretic models within the framework of Becchi-Rouet-Stora-Tyutin (BRST) formalism \cite{17,16,15,24}.

In the BRST formalism, the Bonora-Tonin (BT) superfield approach \cite{bt,bpt} enables us to derive off-shell nilpotent and absolutely anticommuting (anti-)BRST symmetries for a given $p$-form gauge theory (see, e.g. \cite{20,21,25}). The BT formalism has been extended in the regime of SUSY theories (known as supervariable approach) to derive the nilpotent symmetries \cite{22,4,23}. In the supervariable approach, the off-shell nilpotent symmetries have been 
procured by making use of (anti-)chiral supervariables and SUSY invariant restrictions (SUSYIRs).

In the present endeavor, our prime motive is to extend the study of \cite{9} to a general SUSY system. For this, first we superimpose the harmonic oscillator superpotential with other types of superpotentials (such as free particle and `Coulomb-type') and explore the continuous and discrete symmetries of these systems. We then extend our analysis to a generalized form of superpotential from where any superposition of `polynomial-type' superpotentials
can be retrieved.
Another key motive of the present work is to provide a proof for the conjecture (cf. \cite{10} for details) which states that there might exist more than one set of discrete symmetry transformations which provide an analogue for the Hodge duality operator of differential geometry. Furthermore, the idea of supervariable formalism has been mostly exploited to derive the off-shell nilpotent symmetries existing in the SUSY theory (cf. \cite{4,23}). Thus, the final motive
of our present investigation is to make use of (anti-)chiral supervariable approach to derive the {\it on-shell} nilpotent symmetries for the generalized superpotential.

The contents of the paper is organized as follows. Section 2 gives the preliminaries for the study of on-shell symmetries existing for a generalized form of superpotential. Section 3  provides an account for the continuous and discrete symmetries for the superposition of harmonic oscillator superpotential with free particle as well as with the `Coulomb-type' superpotential. Section 4 deals with the study of continuous and discrete symmetries for the `generalized' superpotential, whereas the cohomological aspects of these discrete symmetries are discussed in section 5. Our section 6 contains the derivation of the above mentioned on-shell nilpotent fermionic continuous symmetries, for  `generalized' superpotential, within the framework of (anti-)chiral supervariable approach. Furthermore, we express the Lagrangian corresponding to `generalized'  superpotential in terms of supervariables and establish an equivalence among the translational generators and fermionic symmetries. Finally, in section 7, we summarize our work.

\section{Preliminaries}

We start with the generalized $\mathcal{N} = 2$ SUSY Lagrangian \cite{5} of the following form 
\begin{eqnarray}
L=\frac{\dot{q}^{2}}{2}-\frac{1}{2}W'^{2}+ \text{i}  \bar{\psi}\dot{\psi}-W''\bar{\psi}\psi,
\label{Lg}
\end{eqnarray}
where $q$ is bosonic coordinate, $\psi (\bar{\psi})$ are fermionic variables and $W$ represents superpotential (prime indicates derivative with respect to the bosonic coordinate and over dot represents the time derivative). The above Lagrangian has following sets of fermionic symmetries $(s_1, s_2)$
	\begin{eqnarray}
	s_{1}q=-\text{i}\psi,\quad s_{1}\psi=0, \quad s_{1}\bar{\psi}=\dot{q}+\text{i}W', \label{s1gc}
	\end{eqnarray}
	\begin{eqnarray}
	s_{2}q=\text{i}\bar{\psi},\quad s_{2}\bar{\psi}=0,\quad s_{2}\psi=-\dot{q}+\text{i}W'. \label{s2gc}
	\end{eqnarray}
	We can easily verify that under the above symmetry transformations $(s_1, s_2)$, the Lagrangian \eqref{Lg} transforms as
	\begin{eqnarray}
	s_{1}L =\dfrac{\text{d}}{\text{d}t}(-W'\psi),  \quad s_{2}L =\dfrac{\text{d}}{\text{d}t}(\text{i}\dot{q}\bar{\psi}).
	\label{sL}
	\end{eqnarray}
	The above mentioned fermionic symmetries are on-shell nilpotent of order two
(i.e., $ s_{1}^{2} = s_{2}^{2} = 0$), which can be proven with the help of following Euler-Lagrange equations of motion arising from Lagrangian \eqref{Lg}, as
\begin{eqnarray}
\ddot{q} = - W'W'' - W'''\bar{\psi}\psi, \quad \dot{\bar{\psi}} = \text{i}W''\bar{\psi}, \quad \dot{\psi} = - \text{i}W''\psi.
\end{eqnarray}
	It is interesting to note that the Lagrangian \eqref{Lg} is also endowed with a bosonic symmetry ($s_{w}$) which
 can be constructed from the above mentioned fermionic symmetries as  (cf.  \eqref{s1gc},\eqref{s2gc})
\begin{eqnarray} \label{s_w_phi}
	s_{w} \Phi \; =\; \left\lbrace s_{1},s_{2}\right\rbrace \Phi \; = \; 2\text{i} \dot\Phi,
	\end{eqnarray}
 where $\Phi$ being any of the $q, \psi, \bar{\psi}$. The explicit form of the symmetry transformations under $s_{w}$ is as follows
	\begin{eqnarray}
    s_{w}q = 2 \text{i}\dot{q},\qquad s_{w}\psi= \text{i}\dot{\psi}+W''\bar{\psi}, \qquad s_{w}\bar{\psi}=\text{i}\dot{\bar{\psi}}-W''\psi.
	\end{eqnarray}
It is worthwhile to mention that the study of symmetries of the SUSY Lagrangian leads to the existence of different possible combinations
of superpotentials \cite{5}. Thus, in our present work, we consider the  generalized form of the superpotential as given below \cite{5}
	\begin{eqnarray}
	W'=\omega q + \mu + (\lambda /q), \label{wp}
	\end{eqnarray}
	where $\omega,\mu$ and $\lambda$ are arbitrary constants.
	It is clear from the above expression that various combinations of the constants  $\omega,\mu$ and $\lambda$ yield
	different kinds of superpotentials. For example, $\lambda=0$ returns the superposition of harmonic
oscillator superpotential with free particle, whereas $\mu=0$ gives the superposition of superpotentials such as harmonic oscillator
with the `Coulomb-like' \cite{5}. We shall consider these two cases in detail in the next section.

Before wrapping up this section, we would like to point out that for a system having superpotential of the form \eqref{wp}
the Euler-Lagrange equations of motion turn out to be
	\begin{eqnarray}
	\ddot{q} &=&-\omega^{2}q+\Big(\frac{\lambda}{q^{2}}-\omega\Big)\mu+\frac{\lambda^{2}}{q^{3}}-\frac{2\lambda}{q^{3}}\bar{\psi}\psi,\nonumber\\
	\dot{\bar{\psi}} &=&\text{i}\Big(\omega-\frac{\lambda}{q^{2}}\Big)\bar{\psi}, \qquad
	\dot{\psi} = -\text{i} \Big(\omega-\frac{\lambda}{q^{2}}\Big)\psi. \label{eqn.oml}
	\end{eqnarray}

	\section{Examples : different cases}
	We consider linear combinations of the `generalized' superpotential (cf. \eqref{wp} above) as they return some familiar systems and we can study their
	inherent symmetries. Thus, we have the following cases:
	
	\subsection{Case I: $\lambda=0$}
	In this case, the resultant superpotential represents the superposition of harmonic oscillator with free particle \cite{5}.
	Therefore, we have
	\begin{eqnarray}
	W' \; = \; \omega q + \mu. \label{p1}
	\end{eqnarray}
	The resultant Lagrangian with the superpotential \eqref{p1} looks like
	\begin{eqnarray}
	L^{(1)}=\frac{\dot{q}^{2}}{2}-\frac{1}{2}\omega^{2}q^{2}+ \text{i} \bar{\psi}\dot{\psi}-\omega\bar{\psi}\psi-\frac{\mu^{2}}{2}-\omega\mu q.
	\label{L_HO+FP}
	\end{eqnarray}
The above Lagrangian may be decomposed either into the harmonic oscillator or the free particle Lagrangian depending on the values of $\omega$ and $\mu$.
The corresponding continuous symmetries $(s_{1}^{(1)},s_{2}^{(1)})$ are of the following form
	\begin{eqnarray}
	s_{1}^{(1)}q ={- \text{i}\psi}, \quad s_{1}^{(1)}\bar{\psi} = \dot{q}+ \text{i}\omega q+ \text{i}\mu,\quad s_{1}^{(1)}\psi =0, \label{s1_w_mu}
	\end{eqnarray}
	\begin{eqnarray}
	s_{2}^{(1)}q ={\text{i}\bar{\psi}}, \quad s_{2}^{(1)}\psi = -\dot{q}+\text{i}\omega q+ \text{i}\mu, \quad s_{2}^{(1)}\bar{\psi} =0. \label{s2_w_mu}
	\end{eqnarray}
These symmetries are on-shell nilpotent and the action of the above stated symmetries on the Lagrangian \eqref{L_HO+FP} gives
	\begin{eqnarray}
	s_{1}^{(1)}L^{(1)} = -\dfrac{\text{d}}{\text{d}t}\bigg( (\omega q+\mu)\psi \bigg), \qquad  s_{2}^{(1)}L^{(1)} = \dfrac{\text{d}}{\text{d}t} \big(\text{i}\dot{q}\bar{\psi}\big),
	\end{eqnarray}
and the corresponding Noether conserved charges for the above symmetries are, respectively
	\begin{eqnarray}
	Q^{(1)} =(-\text{i}\dot{q}+\omega q+\mu)\psi, \qquad \bar{Q}^{(1)} = \bar{\psi}(\text{i}\dot{q}+\omega q+\mu).
	\end{eqnarray}
Furthermore, the following bosonic symmetry $s_{w}^{(1)} = \{ s_1^{(1)}, s_2^{(1)}\}$;
	\begin{eqnarray}
& s_{w}^{(1)}q={2\text{i}\dot{q}},\quad s_{w}^{(1)}\psi=\text{i}\dot{\psi}+\omega \psi,\quad s_{w}^{(1)}\bar{\psi}= \text{i}\dot{\bar{\psi}}-\omega\bar{\psi},&
	\label{sw}
	\end{eqnarray}
transforms the Lagrangian \eqref{L_HO+FP} as:
	\begin{eqnarray}
	s_{w}^{(1)}L^{(1)}= 2\text{i}\dfrac{\text{d}}{\text{d}t}\left(\frac{\dot{q}^{2}}{2}-\frac{\omega^{2}q^{2}}{2}+\frac{{\text{i}\bar{\psi}\dot{\psi}}}{2}-\frac{\omega\bar{\psi}\psi}{2}
	-\omega\mu q\right).
	\end{eqnarray}
It is worthwhile to mention couple of points here; first, the bosonic symmetry $s_{w}^{(1)}$ listed in \eqref{sw} reduces to the
form \eqref{s_w_phi} on the on-shell, i.e. by using equations of motion (cf. \eqref{eqm.s1} below). Second, it
can be easily verified that the Lagrangian \eqref{L_HO+FP} transforms into a total time derivative of itself, modulo a constant factor, under $s_{w}^{(1)}$. The corresponding Noether conserved charge for $s_{w}^{(1)}$ can be given as
	\begin{eqnarray}
	Q_{w}^{(1)}=2{\text{i}}\left(\frac{\dot{q}^{2}}{2}+\frac{\omega^{2}q^{2}}{2}+\omega\bar{\psi}\psi+\omega\mu q\right).
	\end{eqnarray}
	 The equations of motion corresponding to $L^{(1)}$ have the following form
	\begin{eqnarray}
	\ddot{q} = -\omega^{2}q-\omega\mu,
	 \quad \dot{\bar{\psi}} =  \text{i}\omega\bar{\psi},
	\quad \dot{\psi} = -\text{i}\omega\psi. \label{eqm.s1}
	\end{eqnarray}

 In addition to the above mentioned continuous symmetries, we also have following discrete symmetries in the system which leave the
 Lagrangian \eqref{L_HO+FP} quasi-invariant
\begin{eqnarray} \label{dis_q0}
 q\longrightarrow - q, \; \quad t\longrightarrow +t, \;\quad \psi \longrightarrow  \pm \text{i}\bar{\psi}, \;\quad \bar{\psi} \longrightarrow \mp \text{i}\psi,
 \quad\omega \longrightarrow -\omega, \; \quad\mu \longrightarrow + \mu,
\end{eqnarray}
where the above set has only parity symmetry (as $q \rightarrow - q$). Furthermore, we have another set of discrete symmetry transformations that
contains both parity and time-reversal which also leaves the Lagrangian \eqref{L_HO+FP} quasi-invariant
\begin{eqnarray} \label{dis_q1}
 q\longrightarrow - q, \; \quad t\longrightarrow - t, \;\quad\psi \longrightarrow  \pm \text{i}\bar{\psi}, \;\quad \bar{\psi} \longrightarrow \pm \text{i}\psi,
 \quad\omega \longrightarrow +\omega, \; \quad\mu \longrightarrow - \mu.
\end{eqnarray}
Finally, there exist two more sets of discrete symmetries containing time reversal symmetry, one with parity and other without parity symmetry, as follows:
\begin{eqnarray} \label{dis_1}
q\longrightarrow \pm q, \; \quad t\longrightarrow -t, \;\quad \psi \longrightarrow  \pm \bar{\psi}, \;\quad \bar{\psi} \longrightarrow \mp \psi,
\quad\omega \longrightarrow \omega, \; \quad\mu \longrightarrow \pm \mu.
\end{eqnarray}
At this juncture, we would like to mention that among the discrete symmetries listed in \eqref{dis_q0} - \eqref{dis_1}, only two sets have physical realization. We would
comment on this, in detail, in section 5 where we discuss the case for a more general superpotential.
Furthermore, it can be shown that the Hamiltonian $(H^{(1)})$ corresponding to $L^{(1)}$  can be written as
$H^{(1)}=\frac{1}{2}\left\lbrace Q^{(1)},\bar{Q}^{(1)} \right\rbrace$.\\

It is worthwhile to mention that we can directly obtain the symmetries and conserved charges for the two sets of individual superpotentials from these
explicit expressions derived above in the following manner.
	
	(a): When substituting the value of $\mu=0$ in the above set of equations, we retrieve the expressions for the SUSY harmonic oscillator Lagrangian. This can be verified in a straightforward manner.
	
     (b): Whereas when $\omega=0$, this leads to the non-relativistic particle case. In a more general sense, the expressions represent a particle
	under a constant potential of magnitude $\mu$. This can also be individually verified.

	\subsection{Case II: $\mu=0$}
	In this case, the superpotential \eqref{wp}, which is analogous to the superposition of harmonic oscillator with `Coulomb-like'
takes the following form
	\begin{eqnarray}
	 W'=\omega q + \lambda/q,
	\end{eqnarray}
	consequently, the Lagrangian can be given as
	\begin{equation}
	L^{(2)}= \frac{\dot{q}^{2}}{2}-\frac{\lambda^{2}}{2q^{2}}-\frac{\omega^{2}q^{2}}{2}+\text{i}\bar{\psi}\dot{\psi}
	+\frac{\lambda}{q^{2}}\bar{\psi}\psi-\omega\bar{\psi}\psi-\omega\lambda. \label{l2}
	\end{equation}
	It is straightforward to check that the above Lagrangian \eqref{l2} has the following sets of fermionic continuous
	symmetries ($s_{1}^{(2)}$ and $s_{2}^{(2)}$)
	\begin{eqnarray}
	 s_{1}^{(2)}q=-\text{i}\psi, \quad s_{1}^{(2)}\psi=0, \quad s_{1}^{(2)}\bar{\psi}=\dot{q} +\text{i}\omega q + \frac{\text{i}\lambda}{q}, \label{s1_w_lambda} \\
	 s_{2}^{(2)}q=\text{i}\bar{\psi},\quad  s_{2}^{(2)}\bar{\psi}=0, \quad s_{2}^{(2)}\psi = -\dot{q} +\text{i}\omega q + \frac{\text{i}\lambda}{q}, \label{s2_w_lambda}
	\end{eqnarray}
where on-shell nilpotency can be proven by virtue of Euler-Lagrange equations of motion (cf. \eqref{eqm.s2} below). Here, the Lagrangian \eqref{l2} transforms into a total derivative under these fermionic continuous symmetries and takes the following forms, respectively
	\begin{eqnarray}
	s_{1}^{(2)}L^{(2)} = -\dfrac{\text{d}}{\text{d}t}\left(\big(\omega q+\frac{\lambda}{q}\big)\psi\right),
	\qquad s_{2}^{(2)}L^{(2)} = \dfrac{\text{d}}{\text{d}t} \big(\text{i}\dot{q}\bar{\psi}\big).
	\end{eqnarray}
In a similar fashion, as we have done in the last subsection, we can construct the bosonic symmetry
$s_{w}^{(2)} = \left\lbrace s_{1}^{(2)}, s_{2}^{(2)}\right\rbrace$ as follows
	\begin{eqnarray} \label{sw2}
	& s_{w}^{(2)}q=2\text{i}\dot{q},\quad s_{w}^{(2)}\psi=\text{i}\dot{\psi}+\displaystyle \Big(\omega-\frac{\lambda}{q^{2}}\Big)\psi, \quad
   s_{w}^{(2)}\bar{\psi}=\text{i}\dot{\bar{\psi}} -\displaystyle \Big(\omega-\frac{\lambda}{q^{2}}\Big)\bar{\psi}.&
	\end{eqnarray}
	It is easy to check that under the above bosonic symmetry, the Lagrangian \eqref{l2} transforms in the following manner
	\begin{eqnarray}
	s_{w}^{(2)}L^{(2)} = \text{i}\dfrac{\text{d}}{\text{d}t}\left({\dot{q}^{2}} - {\omega^{2}q^{2}} - \frac{\lambda^{2}}{q^{2}} + {\text{i}\bar{\psi}\dot{\psi}}
	- {\omega\bar{\psi}\psi} + \frac{\lambda\bar{\psi}\psi}{q^{2}}\right).
	\end{eqnarray}
It is interesting to note that the symmetries listed in \eqref{sw2} can be recast into the form \eqref{s_w_phi}
with the help of equations of motion (cf. \eqref{eqm_s2} below). In this scenario, the Lagrangian \eqref{l2}
transforms into a total time derivative of itself, modulo a constant factor, under the
bosonic symmetry  $s_w^{(2)}$.  The Noether conserved charges corresponding to the fermionic symmetries  \eqref{s1_w_lambda},  \eqref{s2_w_lambda} are, respectively, found to be
	 \begin{eqnarray}
	 Q^{(2)} = \left(-\text{i}\dot{q}+\omega q +\frac{\lambda}{q}\right)\psi,\qquad
	 \bar{Q}^{(2)} = \bar{\psi}\left(\text{i}\dot{q}+\omega q +\frac{\lambda}{q}\right),
	 \end{eqnarray}
	whereas for the bosonic symmetry (cf. \eqref{sw2}), the conserved charge $Q_{w}^{(2)}$ is given by
	 \begin{eqnarray}
	 Q_{w}^{(2)} &=&2\text{i}\left(\frac{\dot{q}^{2}}{2}+\frac{\omega^{2}q^{2}}{2}+\frac{\lambda^{2}}{2q^{2}}+\omega\bar{\psi}\psi
	 -\frac{\lambda\bar{\psi}\psi}{q^{2}}\right).
	 \end{eqnarray}
The Lagrangian \eqref{l2} has the following Euler-Lagrange equations of motion
	\begin{eqnarray}\label{eqm_s2}
	\ddot{q} = - \omega^{2}q+\frac{\lambda^{2}}{q^{3}}-\frac{2\lambda\bar{\psi}\psi}{q^{3}},\qquad
	\dot{\bar{\psi}} =  \text{i}(\omega-\frac{\lambda}{q^{2}})\bar{\psi},\qquad
	\dot{\psi} = - \text{i}(\omega-\frac{\lambda}{q^{2}})\psi. \label{eqm.s2}
	\end{eqnarray}

It is again straightforward to check that, $H^{(2)}=\frac{1}{2}\left\lbrace {Q}^{(2)},\bar{Q}^{(2)}\right\rbrace$, where $H^{(2)}$
represents the Hamiltonian corresponding to $L^{(2)}$. Besides these continuous symmetries there also exists a set of discrete symmetry
transformations as
\begin{eqnarray} \label{dis2_q0}
 q\longrightarrow - q, \; \quad t\longrightarrow +t, \;\quad \psi \longrightarrow  \pm \text{i}\bar{\psi}, \;\quad \bar{\psi} \longrightarrow \mp \text{i}\psi,
 \quad\omega \longrightarrow -\omega, \; \quad\lambda \longrightarrow  -\lambda.
\end{eqnarray}
This set of discrete symmetry transformations contains only parity symmetry, but no time-reversal symmetry. Moreover, we also have
discrete symmetry transformations which consist of both parity and time-reversal symmetry as follows
\begin{eqnarray} \label{dis2_q1}
 q\longrightarrow - q, \; \quad t\longrightarrow -t, \;\quad \psi \longrightarrow  \pm \text{i}\bar{\psi}, \;\quad \bar{\psi} \longrightarrow \pm \text{i} \psi,
 \quad\omega \longrightarrow \omega, \; \quad\lambda \longrightarrow  \lambda.
\end{eqnarray}
Furthermore, there exist two more sets of discrete symmetry transformations one with parity symmetry and another without it along with time-reversal
symmetry, as
\begin{eqnarray} \label{dis_2}
 q\longrightarrow \pm q, \; \quad t\longrightarrow -t, \;\quad \psi \longrightarrow  \pm \bar{\psi}, \;\quad \bar{\psi} \longrightarrow \mp \psi,
 \quad\omega \longrightarrow \omega, \; \quad\lambda \longrightarrow  \lambda.
\end{eqnarray}
Here all the above listed discrete symmetry transformations \eqref{dis2_q0} - \eqref{dis_2} leave the Lagrangian \eqref{l2} quasi-invariant.
It is worthwhile to mention that
among the above mentioned sets of discrete symmetries, only two sets correspond to be physically significant (similar to the previous case). We shall discuss this aspect in section 5.

    Before ending this subsection we would like to comment that in this combination of superpotentials, we retrieve two individual cases.

 (a): When substituting  $\lambda=0$ in the above set of equations, we retrieve the corresponding expressions for the SUSY harmonic oscillator. This can be verified by individually evaluating fermionic and bosonic symmetries (and their corresponding charges) for a harmonic oscillator superpotential.

(b): While substituting $\omega=0$ in the above set of equations, we retrieve the expressions for a classical potential
	 equivalent to $V(q)=\frac{\lambda^{2}}{q^{2}}$ (cf. \cite{5} for details). This can be again verified by separately evaluating
	 fermionic and bosonic symmetries (and their corresponding charges)  for a `Coulomb-like' superpotential.

\section{Superpotentials of type $\displaystyle {W' =\sum_{j=a}^{b}\beta_{j}q^{j}}$}
In this section, we take a more general case with `generalized' superpotential of the form
\begin{eqnarray}
 W'=\sum_{j=a}^{b}\beta_{j}q^{j} ,
\label{wg}
\end{eqnarray}
where $a,b$ could be positive or negative integers in an `appropriate order' and $\beta_{j}$'s are
 constant coefficients. If $a,b$ are taken in `appropriate order' (i.e. $ a, a+1, a+2,..., b=a+n)$ with $n$ being a positive integer,
 we obtain a series of superpotentials with the powers of $q$ changing in geometric progression. It is straightforward to obtain the superpotential \eqref{wp} from this `generalized' superpotential
with the following choices (as a special case).

Let's choose $a=-1$ and $b=1$, this implies
\begin{eqnarray}
\sum_{j=-1}^{1}\beta_{j}q^{j} &=&\beta_{-1}q^{-1}+\beta_{0}q^{0}+\beta_{1}q^{1},
\end{eqnarray}
with identifying $\beta_{-1}=\lambda, \beta_{0}=\mu$ and $\beta_{1}=\omega$, we obtain the superpotential \eqref{wp} as
\begin{eqnarray}
\sum_{j=-1}^{1}\beta_{j}q^{j} &=& \frac{\lambda}{q}+\mu+\omega q.
\end{eqnarray}
Thus, we start with `generalized' superpotential \eqref{wg}, so the Lagrangian can be given as
	\begin{equation}
	L^{(g)}=\frac{\dot{q}^{2}}{2}-\frac{1}{2}\left(\sum_{j=a}^{b}\beta_{j}q^{j}\right)^{2}+\text{i}\bar{\psi}\dot{\psi}
	-\left(\sum_{j=a}^{b}j\beta_{j}q^{j-1}\right)\bar{\psi}\psi.\label{lg} \\
	\end{equation}
The fermionic symmetries $s_{1}^{(g)}$, $s_{2}^{(g)}$ associated with the above mentioned Lagrangian are
\begin{eqnarray}
s_{1}^{(g)}q =-\text{i}\psi, \quad s_{1}^{(g)}\psi =0, \quad s_{1}^{(g)}\bar{\psi} =\dot{q}+\text{i}\sum_{j=a}^{b}\beta_{j}q^{j}, \label{s1g}\\
s_{2}^{(g)}q =\text{i}\bar{\psi},\quad s_{2}^{(g)}\bar{\psi} =0, \quad s_{2}^{(g)}\psi =-\dot{q}+\text{i}\sum_{j=a}^{b}\beta_{j}q^{j}. \label{s2g}
\end{eqnarray}
	Moreover, apart from the above fermionic symmetries as listed in \eqref{s1g}, \eqref{s2g} we have the following bosonic
	symmetry $s_{w}^{(g)}=\left\lbrace s_{1}^{(g)},s_{2}^{(g)}\right\rbrace$ which can be explicitly given as
	\begin{equation}\label{swg}
	s_{w}^{(g)}q = 2\text{i}\dot{q}, \quad s_{w}^{(g)}\psi = \text{i}\dot{\psi}+\sum_{j=a}^{b}j\beta_{j}q^{j-1}\psi,
	\quad s_{w}^{(g)}\bar{\psi} = \text{i}\dot{\bar{\psi}}-\sum_{j=a}^{b}j\beta_{j}q^{j-1}\bar{\psi}.\\	
	\end{equation}
	It is now straightforward to check that under the above set of symmetry transformations, ($s_{1}^{(g)},s_{2}^{(g)},s_{w}^{(g)}$)
	the Lagrangian \eqref{lg} transforms as follows
	\begin{eqnarray}
	s_{1}^{(g)}L^{(g)} &=&-\dfrac{\text{d}}{\text{d}t}\left(\sum_{j=a}^{b}\big(\beta_{j}q^{j}\big)\psi\right), \qquad
	s_{2}^{(g)}L^{(g)} =\dfrac{\text{d}}{\text{d}t}\left(\text{i}\dot{q}\bar{\psi}\right),\\	
	s_{w}^{(g)}L^{(g)} &=& \text{i}\dfrac{\text{d}}{\text{d}t}\left( \dot{q}^{2} - \big(\sum_{j=a}^{b}\beta_{j}q^{j}\big)^{2}
	+ \text{i}\bar{\psi}\dot{\psi} - \big(\sum_{j=a}^{b}j\beta_{j}q^{j-1}\big)\bar{\psi}\psi\right).
	\end{eqnarray}
As it can be seen that the Lagrangian \eqref{lg} transforms to a total derivative, hence $s_{1}^{(g)},s_{2}^{(g)}$ and $s_{w}^{(g)}$ are symmetries of the theory. It is important to point out that  the Lagrangian \eqref{lg} transforms into the total time derivative
of itself, modulo a constant factor, under the bosonic symmetry  $s_{w}^{(g)}$ on the on-shell, where we make use of the equations of motion (cf. \eqref{eqm.g} below). Therefore, the conserved charges associated with these symmetries $s_{1}^{(g)}, s_{2}^{(g)}, s_{w}^{(g)}$ can be given, respectively
	\begin{eqnarray}
	Q^{(g)} &=& \Big(-\text{i}\dot{q}+\sum_{j=a}^{b}\beta_{j}q^{j}\Big)\psi, \qquad
	\bar{Q}^{(g)} = \bar{\psi} \Big(\text{i}\dot{q}+\sum_{j=a}^{b}\beta_{j}q^{j}\Big),\\
	Q_{w}^{(g)} &=& 2\text{i}\left[\frac{\dot{q}^2}{2}+\frac{1}{2}\left(\sum_{j=a}^{b}\beta_{j}q^{j}\right)^{2}
	+\left(\sum_{j=a}^{b}j\beta_{j}q^{j-1}\right)\bar{\psi}\psi \right].
	\end{eqnarray}
The Euler-Lagrange equations of motion can be deduced from \eqref{lg}, as follows
	\begin{eqnarray} \label{eqm.g}
   & \ddot{q} = - \displaystyle \sum_{j=a}^{b}\big(\beta_{j}q^{j}\big)\displaystyle \sum_{k=a}^{b}\big(k\beta_{k}q^{k-1}\big)-\left[{\sum_{j=a}^{b}}j(j-1)\beta_{j}q^{j-2}\right]\bar{\psi}\psi,&\\ \nonumber
       &\dot{\psi} = -\text{i}\displaystyle{\sum_{j=a}^{b}}\big(j\beta_{j}q^{j-1}\big)\psi,
      \qquad \dot{\bar{\psi}} = \text{i}\displaystyle{\sum_{j=a}^{b}}\big(j\beta_{j}q^{j-1}\big)\bar{\psi}.&
\end{eqnarray}
Before wrapping up this section,  we would like to comment that the symmetries $s_{1}^{(g)}$ and $s_{2}^{(g)}$ are on-shell nilpotent (i.e. $s_{1}^{(g)2}=0,s_{2}^{(g)2}=0$ only on using above equations of motion).

Apart from the above mentioned continuous symmetries \eqref{s1g} - \eqref{swg}, the Lagrangian \eqref{lg} also possesses
following discrete symmetries
\begin{eqnarray} \label{ds1}
& q\longrightarrow - q, \; \quad t\longrightarrow +t, \;\quad \psi \longrightarrow  \pm \text{i}\bar{\psi}, \;\quad \bar{\psi} \longrightarrow \mp \text{i}\psi, \\ \nonumber
&\beta_{2n+1} \longrightarrow -\beta_{2n+1}, \; \beta_{2n} \longrightarrow + \beta_{2n}, \; {\text {where  $n$ is any integer.}} & \nonumber
\end{eqnarray}
The above listed discrete symmetry transformations endow only parity symmetry without any time-reversal symmetry.
Let us now consider another set of discrete symmetry transformations which possesses both parity and time-reversal symmetries, as
\begin{eqnarray} \label{ds3}
& q\longrightarrow - q, \; \quad t\longrightarrow -t, \;\quad \psi \longrightarrow  \pm \text{i}\bar{\psi}, \;\quad \bar{\psi} \longrightarrow \pm \text{i}\psi, \\
&\beta_{2n+1} \longrightarrow \beta_{2n+1}, \; \beta_{2n} \longrightarrow - \beta_{2n}, \; {\text {where  $n$ is any integer.}} & \nonumber
\end{eqnarray}
Moreover, there exist another two sets of discrete symmetry transformations, as
\begin{eqnarray}  \label{ds2}
& q\longrightarrow \pm q, \; \quad t\longrightarrow -t, \;\quad \psi \longrightarrow  \pm \bar{\psi}, \;\quad \bar{\psi} \longrightarrow \mp \psi, \\
&\beta_{2n+1} \longrightarrow \beta_{2n+1}, \; \beta_{2n} \longrightarrow \pm \beta_{2n}, \; {\text {where  $n$ is any integer.}} & \nonumber
\end{eqnarray}
In the above, one set is endowed with parity symmetry and another without it. However
time-reversal symmetry is present in both the sets.
It is easy to verify that under these sets of discrete symmetry transformations \eqref{ds1} - \eqref{ds2}, the
Lagrangian \eqref{lg} remains quasi-invariant.

\section{Cohomological aspects}

The continuous symmetries defined in \eqref{s1g} - \eqref{swg} satisfy the following algebra
\begin{eqnarray} \label{s1_s2_sw}
&s_{1}^{(g) 2} = 0, \qquad s_{2}^{(g)2} = 0, \qquad s_{w}^{(g)} = \{s_{1}^{(g)}, s_{2}^{(g)} \} = s_{1}^{(g)} s_{2}^{(g)} + s_{2}^{(g)} s_{1}^{(g)}, & \\ \nonumber
&s_{w}^{(g)} = (s_{1}^{(g)} + s_{2}^{(g)})^{2}, \qquad [s_{w}^{(g)}, s_{1}^{(g)}] = 0, \qquad [s_{w}^{(g)}, s_{2}^{(g)}] = 0.&
\end{eqnarray}
At this juncture, we would like to mention that the two sets of continuous symmetries listed in the previous section
(cf. \eqref{s1_w_mu}, \eqref{s2_w_mu}, \eqref{sw} and \eqref{s1_w_lambda}, \eqref{s2_w_lambda}, \eqref{sw2})
also satisfy the same algebra as listed in \eqref{s1_s2_sw}. Moreover, the above mentioned algebraic relations are satisfied
only on the on-shell where we make use of equations of motion.

On the other hand, the de Rham cohomological operators $(d, \delta, \Delta)$ of differential geometry satisfy the relations given below \cite{12,13}
\begin{eqnarray} \label{d_del_Del}
&d^{2} = 0, \qquad \delta^{2} = 0, \qquad \Delta = \{d,\delta\} = d\delta+\delta d,& \\ \nonumber
&\Delta = (d+\delta)^{2}, \qquad [\Delta, d] = 0, \qquad [\Delta, \delta] = 0,&
\end{eqnarray}
where, $(\delta) d$ are (co-)exterior derivatives and  $\Delta$ is the Laplacian operator. In addition to the expressions
listed above, the co-exterior derivative is related with exterior derivative through the Hodge duality operator ($\ast$) and it is expressed as
\begin{eqnarray} \label{de_rham}
\delta \; = \; \pm \; \ast \; d \; \ast,
\end{eqnarray}
where the $\pm$ signs in the above expression are determined by the degree of the form and the dimensionality of the manifold \cite{12,13}.
Thus, from \eqref{s1_s2_sw} and \eqref{d_del_Del}, we can see that the algebra obeyed by the symmetry transformations is same as the algebra
obeyed by the de Rham cohomological operators. In order to have {\it perfect} analogy between symmetries and cohomological operators the equation
\eqref{de_rham} should also be satisfied \cite{10}.

Let us explicitly check the validity of relationship \eqref{de_rham} in the context of symmetries \eqref{ds1}, \eqref{ds3} and \eqref{ds2},
connecting continuous nilpotent symmetries ($s_{1}^{(g)}, s_{2}^{(g)}$) with discrete symmetry transformations ($\ast$) in the following fashion
\begin{eqnarray} \label{s1_phi}
s_{1}^{(g)}\phi = \pm \ast s_{2}^{(g)}\ast\phi, \qquad  {\text{where}} \quad \phi = q, \psi, \bar \psi,
\end{eqnarray}
and also the existence of inverse of the above relation, as
\begin{eqnarray} \label{s2_phi}
s_{2}^{(g)}\phi = \mp \ast s_{1}^{(g)}\ast\phi.
\end{eqnarray}
Before checking the validity of \eqref{de_rham}, we would like to comment that generally the two successive operations of discrete symmetry
transformations on any general variable $(\phi)$ of the theory yield
\begin{eqnarray}
\ast(\ast\phi) =\pm\phi.
\end{eqnarray}
As far as the discrete symmetry transformations \eqref{ds1} are concerned, it is found to satisfy
\begin{eqnarray}
\ast(\ast\phi) =+\phi.
\end{eqnarray}
Moreover, it can be checked explicitly that the transformations \eqref{ds1} do not satisfy the relation, $s_{1}^{(g)}\phi
= + \ast s_{2}^{(g)}\ast\phi$ (and also $s_{2}^{(g)}\phi = - \ast s_{1}^{(g)}\ast\phi$). So, this set of discrete symmetry transformations
cannot be considered to be a {\it perfect} analogue of the de Rham cohomological operators \cite{10}.

Now let us consider the discrete symmetry transformations \eqref{ds3}, as they satisfy
\begin{eqnarray} \label{dis_phi}
\ast[\ast\phi_{1}] =+\phi_{1},\qquad
\ast[\ast\phi_{2}] =-\phi_{2},
\end{eqnarray}
where $\phi_{1}$ and $\phi_{2}$ are the bosonic and fermionic variables, respectively. However, the discrete symmetry
transformations \eqref{ds3} do not satisfy the relation $s_{1}^{(g)}\phi_{1}=+ \ast s_{2}^{(g)}\ast\phi_{1},$
(and the inverse relation $s_{2}^{(g)}\phi_{1}= - \ast s_{1}^{(g)}\ast\phi_{1}$, too). Additionally, for the transformations \eqref{ds3},
the relation $s_{1}^{(g)}\phi_{2} = - \ast s_{2}^{(g)}\ast\phi_{2}$ (and its inverse relation $s_{2}^{(g)}\phi_{2} = + \ast s_{1}^{(g)}\ast\phi_{2}$)
is also not obeyed. Thus, we can ignore these discrete symmetry transformations while looking for a {\it perfect} analogy \cite{10}.

Finally, let us concentrate on the discrete symmetries listed in \eqref{ds2} which have special significance. These symmetries
satisfy the relations listed in \eqref{dis_phi}. Now considering a set of discrete symmetry transformations from \eqref{ds2}, the one without parity symmetry, i.e.
\begin{eqnarray} \label{ds2_1}
&q\longrightarrow + q, \; \quad t\longrightarrow -t, \;\quad \psi \longrightarrow  \pm \bar{\psi}, \;\quad \bar{\psi} \longrightarrow \mp \psi, &\\  \nonumber
&\beta_{2n+1} \longrightarrow \beta_{2n+1}, \; \beta_{2n} \longrightarrow + \beta_{2n}, \; {\text {where  $n$ is any integer.}} & \nonumber
\end{eqnarray}
These transformations are found to obey relations analogous to \eqref{de_rham}, as listed below
\begin{eqnarray}
	s_{1}^{(g)}\phi_{1}=\pm \ast s_{2}^{(g)}\ast\phi_{1}, \; \qquad s_{2}^{(g)}\phi_{1}=\mp \ast s_{1}^{(g)}\ast\phi_{1},\nonumber \\
	s_{1}^{(g)}\phi_{2}=\mp \ast s_{2}^{(g)}\ast\phi_{2}, \; \qquad s_{2}^{(g)}\phi_{2}=\pm \ast s_{1}^{(g)}\ast\phi_{2}.
	\label{Hodge}
\end{eqnarray}
 Let us further concentrate on the upper signature of discrete symmetry transformations in \eqref{ds2_1}, i.e.
\begin{eqnarray} \label{ds2_1_u}
&q\longrightarrow + q, \; \quad t\longrightarrow -t, \;\quad \psi \longrightarrow  + \bar{\psi}, \;\quad \bar{\psi} \longrightarrow - \psi, &\\  \nonumber
&\beta_{2n+1} \longrightarrow \beta_{2n+1}, \; \beta_{2n} \longrightarrow + \beta_{2n}, {\text {where  $n$ is any integer.}} & \nonumber
\end{eqnarray}
Here, we observe that the discrete symmetries given in \eqref{ds2_1_u} satisfy the relations as
\begin{eqnarray}
	s_{1}^{(g)}\phi_{1}=+ \ast s_{2}^{(g)}\ast\phi_{1}, \; \qquad s_{2}^{(g)}\phi_{1}= - \ast s_{1}^{(g)}\ast\phi_{1},\nonumber \\
	s_{1}^{(g)}\phi_{2} = - \ast s_{2}^{(g)}\ast\phi_{2}, \; \qquad s_{2}^{(g)}\phi_{2} = + \ast s_{1}^{(g)}\ast\phi_{2}.
	\label{Hodge_1}
\end{eqnarray}
These relations are the appropriate ones, in accordance with \eqref{de_rham} (cf. \cite{10} for details). Thus, these discrete symmetry transformations \eqref{ds2_1_u}
along with continuous symmetries \eqref{s1g}, \eqref{s2g} and \eqref{swg} provide the physical realization
of de Rham cohomological algebra existing in the context of differential geometry. Whereas, the discrete symmetries with lower signature
in \eqref{ds2_1} lead to $s_{1}^{(g)}\phi_{1}=- \ast s_{2}^{(g)}\ast\phi_{1}$ and $s_{1}^{(g)}\phi_{2} = + \ast s_{2}^{(g)}\ast\phi_{2}$
(and also the inverse relations) which are not in accordance with \eqref{de_rham}.

Now, considering the discrete symmetry transformations in \eqref{ds2} with parity inversion, i.e.:
\begin{eqnarray} \label{ds2_2}
&q\longrightarrow - q, \; \quad t\longrightarrow -t, \;\quad \psi \longrightarrow  \pm \bar{\psi}, \;\quad \bar{\psi} \longrightarrow \mp \psi, &\\  \nonumber
&\beta_{2n+1} \longrightarrow \beta_{2n+1}, \; \beta_{2n} \longrightarrow - \beta_{2n}, \; {\text {where  $n$ is any integer.}} & \nonumber
\end{eqnarray}
The above mentioned discrete symmetry transformations also satisfy relations which are analogous to \eqref{de_rham}, as
\begin{eqnarray}
	s_{1}^{(g)}\phi_{1}=\mp \ast s_{2}^{(g)}\ast\phi_{1}, \; \qquad s_{2}^{(g)}\phi_{1}=\pm \ast s_{1}^{(g)}\ast\phi_{1},\nonumber \\
	s_{1}^{(g)}\phi_{2}=\pm \ast s_{2}^{(g)}\ast\phi_{2}, \; \qquad s_{2}^{(g)}\phi_{2}=\mp \ast s_{1}^{(g)}\ast\phi_{2}.
	\end{eqnarray}
Out of two sets of discrete symmetries listed in \eqref{ds2_2}, let us focus on the lower signature, as explicitly given below
\begin{eqnarray} \label{ds2_2_l}
&q\longrightarrow - q, \; \quad t\longrightarrow -t, \;\quad \psi \longrightarrow  - \bar{\psi}, \;\quad \bar{\psi} \longrightarrow + \psi,& \\  \nonumber
&\beta_{2n+1} \longrightarrow \beta_{2n+1}, \; \beta_{2n} \longrightarrow - \beta_{2n}, \; {\text {where  $n$ is any integer.}} &  \nonumber
\end{eqnarray}
This set of discrete symmetry transformations satisfy the same relations as in \eqref{Hodge_1}.
Thus, we conclude that the discrete symmetry transformations \eqref{ds2_2_l} also provide a physical analogue to the Hodge duality $(\ast)$ operator \cite{10}.
Whereas, the discrete symmetries with upper signature in \eqref{ds2_2} guide to the relations
such as $s_{1}^{(g)}\phi_{1} = - \ast s_{2}^{(g)}\ast\phi_{1}$ and $s_{1}^{(g)}\phi_{2} = + \ast s_{2}^{(g)}\ast\phi_{2}$ (and also the inverse
relations), which are not in line with \eqref{de_rham}.

Thus, we conclude that none of the two discrete symmetries \eqref{ds1}, \eqref{ds3} can be considered as {\it perfect} symmetry in the realm
of differential geometry, since they do not pave the way to the correct relations as in \eqref{s1_phi} and \eqref{s2_phi}. Thus, among the eight
sets of discrete symmetry transformations \eqref{ds1}, \eqref{ds3} and \eqref{ds2}, two of them [as explicitly listed in \eqref{ds2_1_u} and \eqref{ds2_2_l}]
have an important physical significance where we obtain a physical realization among the symmetries and the de Rham cohomological operators
at the algebraic level. Taking the above relations \eqref{de_rham}, \eqref{s1_phi} and \eqref{s2_phi} into consideration, we can easily
identify continuous symmetries $s_{2}^{(g)}, s_{1}^{(g)}$ as analogue of $\delta, d$ of differential geometry and the relations \eqref{s1_s2_sw} and
\eqref{d_del_Del} imply that $s_{w}^{(g)}$ stands for $\Delta$, the Laplacian operator. Moreover, the Hodge duality $(\ast)$ operator can be realized
in terms of two sets of discrete symmetries as listed in \eqref{ds2_1_u} and \eqref{ds2_2_l}.

Before wrapping up this section, we would like to comment upon the physically relevant discrete symmetries in the case of
superposition of harmonic oscillator superpotential with free particle and also with `Coulomb-like' superpotential
(i.e. Case I and Case II, respectively). As far as Case I is concerned, the discrete symmetries in \eqref{dis_1} without parity and having
upper signature provide the perfect analogue of differential geometrical operators as they satisfy relation \eqref{Hodge_1}. Whereas the
discrete symmetries with parity and having lower signature in \eqref{dis_1} also obey the relations listed in \eqref{Hodge_1} and hence also
provide the physical analogue. Likewise in Case II, the discrete symmetries in \eqref{dis_2} without parity symmetry having upper
signature and with parity symmetry having lower signature satisfy the relation \eqref{Hodge_1} and hence provide physical analogue for the Hodge duality operator.

\section{(Anti-)chiral supervariable approach: on-shell nilpotent symmetries}
To derive the full set of on-shell nilpotent fermionic symmetries we start with \(\mathcal{N}\) = 2 SUSY QM model with
the `generalized' superpotential (cf. section 4) on $(1,2)$-dimensional supermanifold. Here, the supermanifold is characterized
by the bosonic variable $t$ and the Grassmann variables $\theta, \bar{\theta}$ ($\theta^{2} = 0, \bar{\theta}^{2} = 0,
\theta\bar{\theta} + \bar{\theta}\theta = 0$). We make use of SUSYIRs to derive on-shell nilpotent symmetries.

\subsection{On-shell nilpotent symmetries: anti-chiral supervariables}
In order to obtain nilpotent fermionic continuous symmetries $s_{1}^{(g)}$,
we focus on the anti-chiral super-submanifold (cf. \cite{4} for details). Therefore, we generalize ordinary variables to their corresponding anti-chiral
supervariables in the following fashion
 \begin{eqnarray} \label{ac_1}
	{\cal{Q}}(t,\bar{\theta}) &=& q(t)+\bar{\theta}\Lambda(t), \label{d1} \nonumber\\
	 \Psi(t,\bar{\theta}) &=& \psi(t)+\text{i}\bar{\theta}\Omega_{1}(t), \nonumber\\
	 \bar{\Psi}(t,\bar{\theta}) &=&\bar{\psi}(t)+ \text{i}\bar{\theta}\Omega_{2}(t),
	 \end{eqnarray}
here, $\Lambda$ is a fermionic secondary variable, whereas $\Omega_{1}$ and $\Omega_{2}$ are bosonic secondary variables.
To evaluate $\Lambda(t), \Omega_{1}(t)$ and $\Omega_{2}(t)$ we make use of the following SUSYIRs \cite{4}.
First of all, the invariance of $\psi$ under transformation $s_{1}^{(g)}$ (i.e. $s_{1}^{(g)}\psi =0$) implies the following SUSYIR (cf. \cite{10, 4} for details)
\begin{eqnarray} \label{susy_ac_1}
 \Psi(t,\bar{\theta}) \; = \; \psi(t).
\end{eqnarray}
Thus, using \eqref{ac_1} into \eqref{susy_ac_1}, we obtain
\begin{eqnarray}
 \Omega_{1}(t) \; = \; 0. \label{gpac1}
\end{eqnarray}
Second, it is easy to verify that  $s_{1}^{(g)}\big(q\psi\big) =  0$ and $s_{1}^{(g)}\big(\dot{q}\dot{\psi}\big)  =  0$.
Thus, the invariant quantities $(q\psi)$ and $(\dot{q}\dot{\psi})$ lead to the following SUSYIRs, respectively
\begin{eqnarray}
{\cal{Q}}(t,\bar{\theta})\Psi(t,\bar{\theta}) \; = \; q(t)\psi(t),
\qquad \dot{{\cal{Q}}}(t,\bar{\theta})\dot{\Psi}(t,\bar{\theta}) \; = \; \dot{q}(t)\dot{\psi}(t).
\end{eqnarray}		
Making use of \eqref{gpac1}, we obtain $\Lambda(t)\psi(t) = 0$ and $\dot{\Lambda}(t)\dot{\psi}(t) = 0$. As a result, the obvious choice
 we make is: $\Lambda(t)$ must be proportional to $\psi(t)$, i.e.
\begin{eqnarray}
\Lambda(t) \; = \; -\text{i} \psi(t). \label{gpac2}
\end{eqnarray}
Finally, we focus our attention to the combination of following quantities: $\frac{1}{2}\dot{q}^{2}(t) + \text{i} \bar{\psi}(t)\dot{\psi}(t)
 -  \frac{1}{2}\Big(\displaystyle \sum_{j=a}^{b}\beta_{j}q^{j} (t) \Big)^{2}
- \Big(\displaystyle \sum_{j=a}^{b} j\beta_{j}q^{j-1} (t) \Big) \bar{\psi} (t) \psi (t)
+ \text{i} \Big(\displaystyle \sum_{j=a}^{b} \beta_{j}q^{j} (t) \Big)\dot{q} (t) \; \equiv \; \Upsilon(t)$. Here, $\Upsilon(t)$ is itself an invariant quantity under $s_{1}^{(g)}$ (i.e. $s_{1}^{(g)}\Upsilon(t)=0$ without using equations of motion). Thus,  we have the following  SUSYIR
\begin{eqnarray} \label{susyir_ac}
\Gamma(t,\bar{\theta}) \; = \; \Upsilon(t),
\end{eqnarray}
where, $\Gamma(t,\bar{\theta})$ represents the generalization of $\Upsilon(t)$ on the anti-chiral super-submanifold and can be explicitly expressed as
\begin{eqnarray}
\Gamma(t,\bar{\theta}) &=& \frac{1}{2}\dot{{\cal{Q}}}^{2}(t,\bar{\theta})
+ \text{i} \bar{\Psi}(t,\bar{\theta})\dot{\Psi}(t,\bar{\theta})-\frac{1}{2}\Big(\sum_{j=a}^{b}\beta_{j}{\cal{Q}}^{j}(t,\bar{\theta})\Big)^{2}  \nonumber \\
&-& \Big(\displaystyle \sum_{j=a}^{b} j\beta_{j}{\cal{Q}}^{j-1}(t,\bar{\theta}) \Big) \bar{\Psi}(t,\bar{\theta}) \Psi(t,\bar{\theta})
+ \text{i} \Big( \displaystyle \sum_{j=a}^{b} \beta_{j}{\cal{Q}}^{j}(t,\bar{\theta})\Big)\dot{\cal{Q}}(t,\bar{\theta}).
\end{eqnarray}
Substituting for ${\cal{Q}}(t,\bar\theta)$, $\Psi(t,\bar\theta)$, $\bar{\Psi}(t,\bar\theta)$ from \eqref{ac_1}
and with the help of \eqref{gpac1}, equating both sides of \eqref{susyir_ac}, we obtain the following expression
\begin{eqnarray}
&&\dot{q}\dot{\Lambda} - \Big(\displaystyle \sum_{j=a}^{b} \beta_{j}q^{j}\Big)\Big(\displaystyle \sum_{k=a}^{b} k\beta_{k}q^{k-1}\Big)\Lambda
- \Big(\displaystyle \sum_{j=a}^{b} j(j-1)\beta_{j}q^{j-2}\Big)\Lambda \bar{\psi}\psi + \text{i}\Big(\displaystyle \sum_{j=a}^{b} \beta_{j}q^{j}\Big) \dot\Lambda \nonumber \\
&& + \text{i} \Big(\displaystyle \sum_{j=a}^{b} j\beta_{j}q^{j-1}\Big)\Lambda \dot{q} - \Omega_{2}\Big(\dot{\psi}
+ \text{i}\Big(\displaystyle \sum_{j=a}^{b} j\beta_{j}q^{j-1}\Big)\psi\Big) \; = \; 0.
\end{eqnarray}
Substituting for $\Lambda(t)$ from \eqref{gpac2}, we obtain
\begin{eqnarray}
&&-\text{i}\dot{q}\dot{\psi} + \text{i}\Big(\displaystyle \sum_{j=a}^{b} \beta_{j}q^{j}\Big)\Big(\displaystyle \sum_{k=a}^{b} k\beta_{k}q^{k-1}\Big) \psi
+ \Big(\displaystyle \sum_{j=a}^{b} j\beta_{j}q^{j-1}\Big)\dot{q}\psi \nonumber \\
&&+ \Big(\displaystyle \sum_{j=a}^{b} \beta_{j}q^{j}\Big)\dot{\psi} \; = \; \Omega_{2}\Big(\dot{\psi}
+ \text{i}\Big(\displaystyle \sum_{j=a}^{b} j\beta_{j}q^{j-1}\Big)\psi\Big),
\end{eqnarray}
which yields
\begin{eqnarray}
\Omega_{2}(t) \; = \; -\text{i} \Big(\dot{q} (t) + \text{i} \sum_{j=a}^{b}\beta_{j}q^{j} (t) \Big). \label{gpac3}
\end{eqnarray}
The equations \eqref{gpac1}, \eqref{gpac2} and \eqref{gpac3} lead to the following expressions on the anti-chiral super-submanifold
\begin{eqnarray} \label{gac}
&&{\cal{Q}}(t,\bar{\theta}) = q(t)+\bar{\theta}\big(-\text{i}\psi\big)
\equiv q(t)+\bar{\theta}\big(s_{1}^{(g)}q\big), \nonumber \\
&&\Psi(t,\bar{\theta}) = \psi(t)+\bar{\theta}\big(0\big)
\equiv \psi(t)+\bar{\theta}\big(s_{1}^{(g)}\psi\big), \nonumber \\
&&\bar{\Psi}(t,\bar{\theta}) = \bar{\psi}(t)+\bar{\theta}\Big(\dot{q}+\text{i}\sum_{j=a}^{b}\beta_{j}q^{j}\Big)
\equiv \bar{\psi}(t)+\bar{\theta}\big(s_{1}^{(g)}\bar{\psi}\big).
\end{eqnarray}
While examining \eqref{gac}, it is clear that if we perform the translation along the Grassmannian direction $\bar{\theta}$, it gives the same result as $s_{1}^{(g)}$ acting on the corresponding ordinary variable. Thus, the equivalence between
the translational generator $(\partial_{\bar{\theta}})$ and $s_{1}^{(g)}$ can be made as:
$\displaystyle \frac{\partial{}}{\partial{\bar{\theta}}}\Phi(t,\bar{\theta}) = s_{1}^{(g)}\phi(t)$.
Here $\Phi(t,\bar{\theta})$ is the general supervariable defined on the $(1,1)$-dimensional super-submanifold parametrized by $(t, \bar{\theta})$
and $\phi(t)$ is the corresponding ordinary variable \cite{4}.

\subsection{On-shell nilpotent symmetries: chiral supervariables}

To derive the fermionic continuous symmetries $s_{2}^{(g)}$, we move to the chiral super-submanifold characterized by $(t, \theta)$.
The ordinary variables are generalized to their corresponding chiral supervariables in terms of secondary variables as (cf. \cite{4})
\begin{eqnarray} \label{c_1}
	 {\cal{Q}}(t,\theta) &=& q(t)+\theta \tilde{\Lambda}(t),\label{d2}\nonumber\\
	 \Psi(t,\theta) &= &\psi(t)+\text{i}\theta \tilde{\Omega}_{1}(t),\nonumber\\
	 \bar{\Psi}(t,\theta) &=&\bar{\psi}(t)+ \text{i}{\theta}\tilde{\Omega}_{2}(t),
	  \end{eqnarray}
  where $\tilde{\Lambda}$ is a fermionic secondary variable and $\tilde{\Omega}_{1}, \tilde{\Omega}_{2}$ are bosonic variables.
  We consider following SUSYIRs to evaluate $\tilde{\Lambda}(t), \tilde{\Omega}_{1}(t)$ and $\tilde{\Omega}_{2}(t)$;
first, the invariance of $\bar{\psi}$ under $s_{2}^{(g)}$ (i.e. $s_{2}^{(g)}\bar{\psi} = 0$) leads to the SUSYIR $\bar{\Psi}(t,{\theta}) =\bar{\psi}(t)$.
Thus, with the help of \eqref{c_1}, we have
\begin{eqnarray}
\tilde{\Omega}_{2}(t) = 0. \label{gpc1}
\end{eqnarray}
Second, we note that the quantities $\big(q\bar{\psi}\big)$ and $\big(\dot{q}\dot{\bar{\psi}}\big)$ remain invariant under $s_2^{(g)}$
(i.e. $s_{2}^{(g)}\big(q\bar{\psi}\big) = 0$ and $s_{2}^{(g)}\big(\dot{q}\dot{\bar{\psi}}\big) = 0$) which lead to the following SUSYIRs, respectively
\begin{eqnarray}
{\cal{Q}}(t,{\theta})\bar{\Psi}(t,{\theta}) = q(t)\bar{\psi}(t),
\qquad \dot{{\cal{Q}}}(t,{\theta})\dot{\bar{\Psi}}(t,{\theta}) = \dot{q}(t)\dot{\bar{\psi}}(t). \label{81}
\end{eqnarray}
Since $\tilde{\Omega}_{2}(t)=0$ from \eqref{gpc1}, thus using it in \eqref{81} along with \eqref{c_1}, we obtain $\tilde{\Lambda}(t)\bar{\psi}(t) = 0$
and $\dot{\tilde{\Lambda}}(t)\dot{\bar{\psi}}(t) = 0$. So, we conclude that $\tilde{\Lambda}(t)$ must be
proportional to $\bar{\psi}(t)$ and we appropriately choose
\begin{eqnarray}
\tilde{\Lambda}(t) \; = \; \text{i} \bar{\psi}(t). \label{gpc2}
\end{eqnarray}
Finally, we have the following combination $\frac{1}{2}\dot{q}^{2}(t)-\text{i}\dot{\bar{\psi}}(t){\psi}(t)
- \frac{1}{2}\Big(\displaystyle \sum_{j=a}^{b}\beta_{j}q^{j} (t)\Big)^{2} - \Big(\displaystyle \sum_{j=a}^{b} j\beta_{j}q^{j-1} (t)\Big)
\bar{\psi} (t)\psi(t) - \text{i} \Big(\displaystyle \sum_{j=a}^{b} \beta_{j}q^{j} (t) \Big)\dot{q} (t)
\equiv \tilde{\Upsilon}(t)$. Here, $\tilde{\Upsilon}(t)$ is itself an invariant quantity under $s_{2}^{(g)}$ (i.e. $s_{2}^{(g)}\tilde{\Upsilon}(t)=0$ without using equations of motion). Thus, we have the following SUSYIR
\begin{eqnarray} \label{susyir_c}
\tilde{\Gamma}(t,\theta) \; = \; \tilde{\Upsilon}(t),
\end{eqnarray}
where $\tilde{\Gamma}(t,\theta)$ represents the generalization of $\tilde{\Upsilon}(t)$ on the chiral super-submanifold and can be explicitly given by
\begin{eqnarray}
\tilde{\Gamma}(t,\theta) &=& \frac{1}{2}\dot{{\cal{Q}}}^{2}(t,{\theta})
-\text{i}\dot{\bar{\Psi}}(t,{\theta}){\Psi}(t,{\theta})-\frac{1}{2}\Big(\sum_{j=a}^{b}\beta_{j}{\cal{Q}}^{j}(t, \theta)\Big)^{2} \nonumber \\
&-& \displaystyle \Big(\sum_{j=a}^{b} j\beta_{j}{\cal{Q}}^{j-1}(t,\theta) \Big) \bar{\Psi}(t,\theta)\Psi(t,\theta)
- \text{i} \Big( \displaystyle \sum_{j=a}^{b} \beta_{j}{\cal{Q}}^{j}(t,\theta)\Big)\dot{\cal{Q}}(t,\theta).
\end{eqnarray}
Making substitutions for ${\cal{Q}}(t,\theta)$, $\Psi(t,\theta)$, $\bar{\Psi}(t,\theta)$ from \eqref{c_1} and with the help of \eqref{gpc1},
equating both  sides of \eqref{susyir_c}, we obtain
\begin{eqnarray}
&& \dot{q}\dot{\tilde{\Lambda}} - \Big(\displaystyle \sum_{j=a}^{b} \beta_{j}q^{j}\Big) \Big(\displaystyle \sum_{k=a}^{b} k\beta_{k}q^{k-1}\Big) \tilde{\Lambda}
- \Big(\displaystyle \sum_{j=a}^{b} j(j-1)\beta_{j}q^{j-2}\Big)\tilde{\Lambda} \bar{\psi}\psi
- \text{i}\Big(\displaystyle \sum_{j=a}^{b} \beta_{j}q^{j}\Big) \dot{\tilde{\Lambda}} \nonumber \\
&&  - \text{i}\Big(\displaystyle \sum_{j=a}^{b} j\beta_{j}q^{j-1}\Big)\tilde{\Lambda} \dot{q} - \tilde{\Omega}_{1}\Big(\dot{\bar{\psi}}
- \text{i}(\displaystyle \sum_{j=a}^{b} j\beta_{j}q^{j-1})\bar{\psi}\Big) \; = \; 0.
\end{eqnarray}
After substituting for $\tilde{\Lambda}(t)$ from \eqref{gpc2}, we obtain
\begin{eqnarray}
&&\text{i}\dot{q}\dot{\bar{\psi}} - \text{i}\Big(\displaystyle \sum_{j=a}^{b} \beta_{j}q^{j}\Big) \Big(\displaystyle \sum_{k=a}^{b} k\beta_{k}q^{k-1}\Big)\bar{\psi}
+ \Big(\displaystyle \sum_{j=a}^{b} j\beta_{j}q^{j-1}\Big)\dot{q}\bar{\psi} \nonumber \\
&&+ \Big(\displaystyle \sum_{j=a}^{b} \beta_{j}q^{j}\Big)\dot{\bar{\psi}} \; = \; \tilde{\Omega}_{1}\Big(\dot{\bar{\psi}}
- \text{i}\Big(\displaystyle \sum_{j=a}^{b} j\beta_{j}q^{j-1}\Big)\bar{\psi}\Big),
\end{eqnarray}
which finally yields
\begin{eqnarray}
 \tilde{\Omega}_{1}(t)=-\text{i}\Big(-\dot{q} (t) + \text{i} \sum_{j=a}^{b}\beta_{j}q^{j} (t) \Big). \label{gpc3}
\end{eqnarray}
Thus, the obtained expressions for $\tilde{\Lambda} (t), \tilde{\Omega}_{1} (t)$ and $\tilde{\Omega}_{2} (t)$ enable us to write expansion of
supervariables in chiral super-submanifold as follows
\begin{eqnarray} \label{gc}
&&{\cal{Q}}(t,\theta) = q(t)+\theta\big(\text{i}\bar{\psi}\big)
\equiv q(t)+\theta\big(s_{2}^{(g)}q\big),  \nonumber \\
&&\Psi(t,\theta) = \psi(t)+\theta\Big(-\dot{q}+\text{i}\sum_{j=a}^{b}\beta_{j}q^{j}\Big)
\equiv \psi(t)+\theta\big(s_{2}^{(g)}\psi\big),  \nonumber \\
&&\bar{\Psi}(t,\theta) = \bar{\psi}(t)+\theta(0)
\equiv \bar{\psi}(t)+\theta\big(s_{2}^{(g)}\bar{\psi}\big).
\end{eqnarray}

Before wrapping up this subsection, we would like to comment that it is evident from \eqref{gc} that
$\displaystyle \frac{\partial}{\partial{\theta}}\Phi(t,\theta) = s_{2}^{(g)}\phi(t)$. Here $\Phi(t,\theta)$ is the general supervariable
in the super-submanifold characterized by $(t,\theta)$ and $\phi(t)$ is the corresponding ordinary variable.

\subsection{Invariance of Lagrangian: supervariable approach}
We can generalize the starting Lagrangian \eqref{lg} onto the $(1,1)$-dimensional (anti-)chiral super-submanifold in the following manner
\begin{eqnarray}
	\tilde{L}_{0}^{(ac)}&=&\frac{1}{2}\dot{{\cal{Q}}}^{(1)}(t,\bar{\theta})\dot{{\cal{Q}}}^{(1)}(t,\bar{\theta})
	+\text{i}\bar{\Psi}^{(1)}(t,\bar{\theta})\dot{\Psi}^{(1)}(t,\bar{\theta})-\frac{1}{2}\tilde{W}'({\cal{Q}}^{(1)})\tilde{W}'({\cal{Q}}^{(1)})\nonumber\\
	&-&\tilde{W}''({\cal{Q}}^{(1)})\bar{\Psi}^{(1)}(t,\bar{\theta})\Psi^{(1)}(t,\bar{\theta}),
	\label{lac}
	\end{eqnarray}
	and
	\begin{eqnarray}
	\tilde{L}_{0}^{(c)}&=&\frac{1}{2}\dot{{\cal{Q}}}^{(2)}(t,{\theta})\dot{{\cal{Q}}}^{(2)}(t,{\theta})+\text{i} \bar{\Psi}^{(2)}(t,{\theta})\dot{\Psi}^{(2)}(t,{\theta})
	-\frac{1}{2}\tilde{W}'({\cal{Q}}^{(2)})\tilde{W}'({\cal{Q}}^{(2)})\nonumber\\
	&-&\tilde{W}''({\cal{Q}}^{(2)})\bar{\Psi}^{(2)}(t,{\theta})\Psi^{(2)}(t,{\theta}).
	\label{lc}
	\end{eqnarray}
	Here, the superscripts $(1)2$ denote the expansion of the supervariables in the (anti-)chiral directions and the superscripts $(ac)c$
	denote the generalization of Lagrangian along the (anti-)chiral directions. The Taylor expansion for the superpotential
	derivatives $\tilde{W}'$ and $\tilde{W}''$ for the anti-chiral case are given as
	\begin{eqnarray} \label{W_ac}
	\tilde{W}'({\cal{Q}}^{(1)}) &=& W'(q)- \text{i}\bar{\theta}W''(q)\psi(t),\nonumber\\
	\tilde{W}''({\cal{Q}}^{(1)}) &=& W''(q)- \text{i} \bar{\theta}W'''(q)\psi(t).
	\end{eqnarray}
	And for the chiral case, we have
	\begin{eqnarray} \label{W_c}
	\tilde{W}'({\cal{Q}}^{(2)}) &=& W'(q)+\text{i}{\theta}W''(q)\bar{\psi}(t),\nonumber\\
	\tilde{W}''({\cal{Q}}^{(2)}) &=& W''(q)+\text{i}{\theta}W'''(q)\bar{\psi}(t).
	\end{eqnarray}
	Now, substituting the expressions \eqref{gac}, \eqref{W_ac} into $\tilde{L}_{0}^{(ac)}$ and \eqref{gc}, \eqref{W_c} into $\tilde{L}_{0}^{(c)}$,
	we obtain the following expressions
	\begin{eqnarray} \label{L_ac}
	\tilde{L}_{0}^{(ac)} &=&\frac{\dot{q}^{2}}{2}-\frac{1}{2}\Big(\sum_{j=a}^{b}\beta_{j}q^{j}\Big)^{2}+ \text{i} \bar{\psi}\dot{\psi}
	-\Big(\sum_{j=a}^{b}j\beta_{j}q^{j-1}\Big)\bar{\psi}\psi-\bar{\theta}\dfrac{\text{d}}{\text{d}t}\Big(\sum_{j=a}^{b}\beta_{j}q^{j}\psi\Big),
\end{eqnarray}
\begin{eqnarray} \label{L_c}
	\tilde{L}_{0}^{(c)} &=&\frac{\dot{q}^{2}}{2}-\frac{1}{2}\Big(\sum_{j=a}^{b}\beta_{j}q^{j}\Big)^{2}+ \text{i} \bar{\psi}\dot{\psi}
	-\Big(\sum_{j=a}^{b}j\beta_{j}q^{j-1}\Big)\bar{\psi}\psi+\text{i}{\theta}\dfrac{\text{d}}{\text{d}t}\Big(\dot{q}\bar{\psi}\Big).
	\end{eqnarray}
	It is evident from the above expressions that, taking the translational derivative of \eqref{L_ac} and \eqref{L_c} along the
	$\bar{\theta}, \theta$ direction, respectively we obtain
	\begin{eqnarray}
	\dfrac{\partial}{\partial\bar{\theta}}\tilde{L}_{0}^{(ac)}\; =\;-\dfrac{\text{d}}{\text{d}t}\Big( \sum_{j=a}^{b}\beta_{j}q^{j}\psi \Big),
	\quad \dfrac{\partial}{\partial\theta}\tilde{L}_{0}^{(c)}\;=\; \dfrac{\text{d}}{\text{d}t}\big(\text{i}\dot{q}\bar{\psi}\big).
	\end{eqnarray}
Thus, the translation along the anti-chiral direction ($\bar{\theta}$) on the superspace Lagrangian \eqref{lac} gives the same result
as $s_{1}^{(g)}$ acting on the SUSY Lagrangian \eqref{lg}. Similarly, the translation of superspace Lagrangian \eqref{lc} along the
chiral direction ($\theta$) also provides the same result as $s_{2}^{(g)}$ acting on the SUSY Lagrangian \eqref{lg}. From these observations,
we can draw the equivalences $\dfrac{\partial}{\partial\bar{\theta}}\equiv s_{1}^{(g)}$ and $\dfrac{\partial}{\partial\theta}\equiv s_{2}^{(g)}$.
Furthermore, these translational generators ($\partial_{\bar{\theta}}, \partial_{\theta}$) are nilpotent (cf. \cite{4}).
These translational generators along the (anti-)chiral
directions in super-submanifold provide the proof for nilpotency of fermionic symmetries in a straightforward manner.

\section{Conclusions}

In our present investigation, we have extensively studied about the genre of symmetries existing in the coupled
\(\mathcal{N}\) = 2 SUSY QM models. In particular, we have examined the superposition of harmonic oscillator superpotential 
with free particle and `Coulomb-type' superpotentials. Moreover, we have considered a `generalized' superpotential which, in turn, 
contains the previous two examples as a limiting case (cf. section 4 for details). We have
inferred that all the above mentioned  coupled systems are endowed with  two sets of on-shell nilpotent, fermionic continuous symmetries as well as a set of bosonic symmetry transformations. The corresponding conserved charges
have also been derived.

Further we have shown that, in each case, there exist eight sets of discrete symmetry transformations that leave the corresponding
Lagrangian quasi-invariant. However, among these eight sets of discrete symmetries only two of them (in each case) provide a physical realization of the Hodge duality operator. 
It is clear from \eqref{s1_s2_sw} and \eqref{d_del_Del} that the algebra satisfied by the fermionic and
bosonic symmetries are same as that of the algebra existing among the de Rham cohomological operators $(d, \delta, \Delta)$.
Furthermore, on a given manifold the exterior $(d)$ and co-exterior $(\delta)$ derivatives are related with one another through
the Hodge duality operator $(\ast)$ see, \eqref{de_rham}. We have shown that this Hodge duality operator finds its physical analogue
in terms of two sets of discrete symmetries of the theory, e.g. \eqref{ds2_1_u} and \eqref{ds2_2_l}.
Thus, we have been able to provide a proof of the conjecture (by considering three different examples) which endorses the existence
of more than one set of discrete symmetry transformations as the analogue of Hodge duality operation \cite{10}.

Moreover, we have applied the supervariable approach to \(\mathcal{N}\) = 2 SUSY QM model with a `generalized' superpotential.
For this, the ordinary variables of the theory have been generalized to the corresponding supervariables onto  the
$(1,1)$-dimensional (anti-)chiral super-submanifold. The on-shell nilpotent
fermionic symmetries  existing in the above mentioned model have been derived by making use of SUSYIRs.
The form of SUSYIRs employed here gives a general norm for the invariant restrictions for any kind of
superposition of potentials falling under the category discussed in the present investigation.
Furthermore, we have generalized the starting Lagrangian \eqref{lg} onto the (anti-)chiral super-submanifold and with the help of that
we have provided an equivalence among fermionic symmetries and translational generators. Finally, the nilpotency of
the fermionic symmetries has been captured in a straightforward manner with the aid of nilpotent translational generators. \\

\noindent
{\bf Acknowledgments} The support from FRG scheme of National Institute of Technology Calicut is thankfully acknowledged.
One of the authors (SG) would like to thank R. Kumar for fruitful discussions. \\

\bibliographystyle{apalike}

\end{document}